\documentstyle [prl,twocolumn,aps] {revtex}
\begin{document}

\title{NON-EXTENSIVE THERMOSTATISTICAL APPROACH OF THE PECULIAR 
VELOCITY FUNCTION OF GALAXY CLUSTERS}

\author{A. Lavagno$^{1,3}$, G. Kaniadakis$^{1,3}$, 
M. Rego-Monteiro$^{2,3}$, P. Quarati$^{1,4}$ and C. Tsallis$^2$}

\address{$^1$ Dipartimento di Fisica-INFM, Politecnico di Torino, 
C.so Duca degli Abruzzi 24, I-10129, Italy \\
$^2$ Centro Brasileiro de Pesquisas Fisicas, 
Rua Xavier Sigaud 150, 22290-180 Rio de Janeiro-RJ, Brazil\\
$^3$ INFN, Sezione di Torino, Italy \\
$^4$ INFN, Sezione di Cagliari, Italy}

\maketitle

\begin{abstract}
We show that the observational data recently provided by Giovanelli et al. 
(1996 a, b) and discussed by Bahcall and Oh (1996) concerning the 
velocity distribution of clusters of galaxies can be naturally fitted by a 
statistical distribution which generalizes the Maxwell-Boltzmann one 
(herein recovered for the entropic index $q=1$). Indeed a recent generalization 
of the Boltzmann-Gibbs thermostatistics suggests for this problem that the 
probability function is given, within a simple phenomenological model, by
\begin{eqnarray}
P(>v) & = & \frac{ \int_{v}^{v_{max}} dv 
\left[ 1-(1-q)(v/v_0)^2 \right]^\frac{q}{1-q} } {\int_{0}^{v_{max}} dv 
\left[ 1-(1-q)(v/v_0)^2 \right]^\frac{q}{1-q}}, \nonumber \\
v_{max} & \equiv & \left\{
\begin{array}{ll}
v_0 (1-q)^{-1/2} & \mbox{if $q<1$}   \\
\infty & \mbox{if $q \geq 1$.} 
\end{array}
\right. \nonumber
\end{eqnarray}
A remarkably good fitting with the data is obtained for 
$q=0.23^{+0.07}_{-0.05}$ and 
$v_0=490\pm 5$ km s$^{-1}$.
\end{abstract}

\vspace{1cm}

The motions of clusters and group of galaxies have been studied by 
Bahcall et al. (1994). The sample of clusters of galaxies 
velocities 
recently observed by Giovanelli et al. 
(1996 a, b) based on Tully-Fisher distances of Sc galaxies has been examined 
and compared with model expectations by Bahcall and Oh (1996).
In contrast with previous analysis of the data, the actual 
observed velocity function does not show a tail of high velocity 
clusters (clusters with velocities greater than $\sim 600$ km s$^{-1}$ can 
be found with a probability less than five per cent).
Bahcall and Oh (1996) determine the cluster velocity function 
using simulations based on four different cosmological models: \\
(a) CDM (cold dark matter), $\Omega \equiv$ {\it dimensionless matter 
density} $= 0.3$; \\
(b) PBI (primeval baryonic isocurvature), $\Omega = 0.3$; \\
(c) CDM, $\Omega = 1.0$; \\
(d) HDM (hot dark matter), $\Omega = 1.0$. \\
These authors conclude that 
model (a) and marginally model (b), in spite of the high energy tail, 
are consistent with the observed data (see Fig. 1), while models (c) 
and (d) show a too large high velocity tail to be in agreement with data.
However, if we look in detail at the observations and simulations curves, 
we can see that not even the CDM, $\Omega = 0.3$ model is completely 
satisfactory, particularly in the region of  velocities larger than 
$350$ km s$^{-1}$. 
In fact, none of the four theoretical curves is fully satisfactory.

We want to show in this letter that to understand the 
Giovanelli et al. (1996 a, b) 
observational data is not 
simply a problem of which cosmological model is used, but it  
also depends on which statistical mechanics is used. In fact, 
if we adopt the recently introduced non-extensive statistics 
(Tsallis 1988, Curado and Tsallis 1991) 
to calculate the probability distribution function of 
cluster peculiar velocities, the agreement with 
experimental results is very satisfactory, particularly in foreseing that 
the number of galaxy clusters with velocities greater than 
$\sim 600$ km s$^{-1}$ is practically zero.
The non-extensive statistics, among several other applications, has been 
recently applied in calculating matter distribution of self-gravitating 
systems (Plastino and Plastino 1993, Aly 1993, 
Boghosian 1996), 
turbulence (Boghosian 1996), 
Levy- and correlated-like anomalous 
diffusions (Zanette and Alemany 1995, 
Tsallis {\it et al.} 1995, Tsallis and Bukman 1996), 
solar neutrino fluxes (Kaniadakis {\it et al.} 1996), 
a cosmological model (Hamity and Barraco 1996), 
as well as in developing a dynamic 
linear response theory (Rajagopal 1996).

We firstly give a brief account of the non-extensive statistics. 
Then we calculate by means of it the probability of finding galaxy clusters 
of certain velocity $v$. Results are shown in the two Figures and discussed; 
conclusions are outlined.

In order to cover non-extensive systems (long-range microscopy memory, 
long range forces, fractal space time) the following generalized entropy 
has been proposed: 
\begin{equation}
\label{1}
S_q = k \frac{1-\sum_{i}p_{i}^{q}}{q-1} \ \ \ \ \ \ \  \left( \sum_{i} p_i = 1; 
\ \ \ \ \  q \epsilon \Re \right) ,
\end{equation}
where $k$ is a positive constant.   
Optimization of $S_q$ yields, for the canonical ensemble,
\begin{eqnarray}
\label{2}
\rho_i  & = & Z_q^{-1} \left[ 1-(1-q) \beta \epsilon_i 
\right]^\frac{1}{1-q}, \\
\label{3}
Z_q     & \equiv & \sum_{i} \left[ 1-(1-q) \beta \epsilon_i 
\right]^\frac{1}{1-q} 
\end{eqnarray}
and, when $q \rightarrow 1$, the Boltzmann-Gibbs result 
($\rho_i \propto e^{-\beta \epsilon_i}$) is recovered. 
Let us stress that, for $q<1$, $\rho_i$ vanishes for 
$\left[ 1-(1-q) \beta \epsilon_i \right] \leq 0$. 
Within this formalism the observational data are to be identified with 
$\langle A \rangle_q = \sum_{i} \rho_{i}^{q} A_i$ where $\left\{ A_i \right\}$ 
are the eigenvalues of an arbitrary observable $A$.

The application of this thermostatistics to stellar systems has 
shown that sensible distribution functions (stellar polytropes) can be 
derived if $q \in (-\infty, 7/9)$. Distributions with $q<1$ give rise 
to the spatial cutoff of the mass distribution and the finite mass 
of the stellar polytrope (Plastino and Plastino 1993, 
Aly 1993 and Boghosian 1996).
Observed distributions of pure-electron plasma with two dimensional turbulence 
are well described with $q=0.5$ (Boghosian 1996).
The logistic map at its threshold to chaos can be described using $q=0.24$ 
(Tsallis {\it et al.} 1996).
Solar neutrino fluxes can be understood using non-extensive statistics for the 
central core plasma and values of $q$ slightly below one: $q=0.997$ 
(ion plasma ) and $q=0.976$ (electron plasma) 
(Kaniadakis et al. 1996, Quarati et al. 1996).

We calculate now the probability $P(>v)$ of finding 
one-dimensional peculiar velocity function of clusters of galaxies (relative to 
a comoving cosmic frame) greater than $v$. It is given by 
\begin{eqnarray}
\label{4}
P(>v) & = & \frac{ \int_{v}^{v_{max}} dv 
\left[ 1-(1-q)(v/v_0)^2 \right]^\frac{q}{1-q} } {\int_{0}^{v_{max}} dv 
\left[ 1-(1-q)(v/v_0)^2 \right]^\frac{q}{1-q}},  \\
\label{5}
v_{max} & \equiv & \left\{
\begin{array}{ll}
v_0 (1-q)^{-1/2} & \mbox{if $q<1$},   \\
\infty & \mbox{if $q \geq 1$}, 
\end{array}
\right. 
\end{eqnarray}
where $v_0$ is a phenomenological characteristic velocity which 
incorporates all the averages over cluster configurations taking into 
account their gravitational interactions.

We find the best fit of the observed data and report it on Fig.1. 
The entropic parameter is $q= 0.23^{+0.07}_{-0.05}$ and 
$v_0 = 490 \pm 5$
 km s$^{-1}$. 
This value of $q$, very different from one, reflects the fact 
that the long-range gravitational forces are, of course, 
in the case of galaxy clusters, very important. 
Moreover, by means of the relation 
$n \equiv polytropic\;index\;=3/2 + q/(1-q)$ 
(Boghosian 1996), we may derive that 
the system is characterized by $n \approx 7/4$. The fact that this value lies 
in the range of the 
typically observed (and theoretically allowed: Plastino and Plastino 1993, 
Aly 1993, Boghosian 1996) values constitutes a further, 
and independent, confirmation of the possible 
relevance of the present proposal. \\
Peculiar of the non-extensive statistics is that our 
distribution is cut at $v \geq 520$ km/s, in perfect consistence with the 
observed data.
In Fig. 2 we show $P(>v)$ of Eq.(\ref{4}) for six different values of 
$q$: $q=1$ (implicitely adopted by Bahcall and Oh 1996) 
is the Maxwell-Boltzmann distribution, $q=0$ is a straight line, 
the $q<0$ curves show clearly a pronounced cut tail.

Our results indicate that in order to distinguish which, among the four models 
judiciously used by 
Bahcall and Oh (1996) (each of them using several fitting 
parameters), is the most appropriate to describe the velocity 
distribution, it would be necessary to repeat the simulations within the 
non-extensive statistics with $q= 0.23^{+0.07}_{-0.05}$. 
The best value of $v_0$ can be selected and, possibly, one could obtain 
information on the quantity $\Omega$.
Finally, we note that the efficiency coming from the modification of 
the statistics is greater and more important than the efficiency obtained 
from the modification of the model (precisely the same occured 
for experiments done in pure-electron plasma
turbulence: see Huang and Driscoll 1994 and Boghosian 1996). 
In fact, our fit of the experimental 
results, based on the non-extensive statistics, has been derived within 
a classical ideal gas picture. Incidentally, we notice that the value 
$q= 0.23^{+0.07}_{-0.05}$  
is very close to the value $q=0.24$, which is appropriate to chaotic dynamics
(Tsallis et al. 1996). It would be interesting to 
investigate if this approximate equality 
between the two above figures is just a numerical coincidence or is due to some 
fundamental reasons.

One of us (C.T.) gratefully acknowledges warm hospitality received at the 
Politecnico - Torino and INFN - Cagliari, as well as useful remarks 
by A.R. Plastino.

\centerline{\bf Figure Captions}
\noindent
Fig. 1. -- The observed data of the clusters velocity function are 
reported together with
the four curves corresponding to the four different models of $P(>v)$ 
elaborated by Bahcall and Oh (see their Fig. 1b). 
Our fit of the observed CVF, based on 
non-extensive statistics ($q= 0.23$ and $v_0=490$ km s$^{-1}$), 
is also shown. \\
Fig. 2. -- The function $P(>x)$ versus $x=v/v_0$ is reported for six 
different values of 
the entropic parameter $q$  $(-5,-1,0,0.5,1,2)$. The case $q=1$ is the well 
known Maxwell-Boltzmann distribution. We note that the curve $q=0$ is a 
straight line.  Curves with negative values of $q$ have an abruptly cutted tail.

\end{document}